\documentclass[
 reprint,
superscriptaddress,
 amsmath,amssymb,
 aps,
]{revtex4-1}

\usepackage{graphicx}
\usepackage{hyperref}
\usepackage{xcolor,soul}
\usepackage{amsmath}

\begin{document}
\preprint{GU--08--20}

\title{Rapidly Spinning Compact Stars with Deconfinement Phase Transition}

\author{Tuna Demircik}
\email{tuna.demircik@apctp.org}
\affiliation{Department of Physics, Ben-Gurion University of the Negev, Beer-Sheva 84105, Israel}
\affiliation{Asia Pacific Center for Theoretical Physics, Pohang, 37673, Korea}
\author{Christian Ecker}
\email{ecker@itp.uni-franfurt.de}
\affiliation{Institut f\"ur Theoretische Physik, Goethe Universit\"at, Max-von-Laue-Str. 1, 60438 Frankfurt am Main, Germany}
\author{Matti J\"arvinen}
\email{matti.jarvinen@apctp.org}
\affiliation{The Raymond and Beverly Sackler School of Physics and Astronomy, Tel Aviv University, Ramat Aviv 69978, Tel Aviv, Israel}
\affiliation{Asia Pacific Center for Theoretical Physics, Pohang, 37673, Korea}

\date{\today}

\begin{abstract}
We study rapidly spinning compact stars with equations of state featuring a first order phase transition between strongly coupled nuclear matter and deconfined quark matter by employing the gauge/gravity duality.
We consider a family of models, which allow purely hadronic uniformly rotating stars with masses up to approximately $2.9\, \mathrm{M}_\odot$, and are therefore compatible with the interpretation that the secondary component ($2.59^{+0.08}_{-0.09}\, \mathrm{M}_\odot$) in GW190814 is a neutron star. 
These stars have central densities several times the nuclear saturation density so that strong coupling and non-perturbative effects become crucial.
We construct models where the maximal mass of static (rotating) stars $M_{\mathrm{TOV}}$ ($M_{\mathrm{max}}$) is either determined by the secular instability or a phase transition induced collapse.
We find largest values for $M_{\mathrm{max}}/M_{\mathrm{TOV}}$ in cases where the phase transition determines $M_{\mathrm{max}}$, which shifts our fit result to $M_{\mathrm{max}}/M_{\mathrm{TOV}} = 1.227^{+0.031}_{-0.016}$, a value slightly above the Breu-Rezzolla bound $1.203^{+0.022}_{-0.022}$ inferred from models without phase transition.
\end{abstract}

\maketitle

\section{Introduction}\label{sec:1}

The LIGO/VIRGO collaboration recently announced the gravitational wave event GW190814 \cite{Abbott:2020khf}. 
This event was identified as the merger of a $23.2 ^{+1.1}_{-1.0}\,\mathrm{M}_\odot$ black hole and a $2.59^{+0.08}_{-0.09}\, \mathrm{M}_\odot$ object making it the detection of a binary merger with the most unequal mass ratio $0.112^{+0.008}_{-0.009}$ so far.
While the primary component is conclusively a black hole, the nature of the lighter companion remains unclear.
It falls into the so-called mass-gap region and is either the lightest black hole or the heaviest neutron star (NS) ever observed in a binary system. 
The secondary component is significantly heavier than the most massive known pulsars, including PSR  J1614-2230  ($1.908^{+0.016}_{-0.016}\, \mathrm{M}_\odot$) \cite{Demorest2010ATN,Arzoumanian:2017puf}, PSR J0348+0432 ($2.01^{+0.04}_{-0.04}\, \mathrm{M}_\odot$) \cite{Antoniadis:2013pzd} and MSP J0740+6620 ($2.14^{+0.1}_{-0.09}\, \mathrm{M}_\odot$) \cite{Cromartie:2019kug}.
The mass of the secondary component is also well above the upper bound on the maximum mass of non-rotating NSs $\approx 2.3\, \mathrm{M}_\odot$ determined from the electromagnetic counterpart of GW170817 \cite{Rezzolla:2017aly,Margalit:2017dij,Shibata:2019ctb}.

The possibility that the lighter companion could be a rapidly rotating NS, or a small black hole of that origin, was pointed out in \cite{Most:2020bba} and used to determine a lower bound on the maximum mass $M_{\mathrm{TOV}}<2.08^{+0.04}_{-0.04}\, \mathrm{M}_\odot$ of non-rotating stars.

There exist already a number of studies addressing the nature of the smaller companion of GW190814 and its compatibility with a rapidly rotating NS \cite{Dexheimer:2020rlp,Godzieba:2020tjn,Lim:2020zvx,Tews:2020ylw,Fattoyev:2020cws,Safarzadeh:2020ntc,Zhang:2020zsc,Roupas:2020jyv,Essick:2020ghc}.
We argue that there is an important strong coupling aspect to this question.
Namely, the maximal mass of rotating ($M_{\mathrm{max}}$) and static ($M_{\mathrm{TOV}}$) NSs is sensitive to the equation of state (EoS) at high density where effective nuclear theory models become unreliable and the gauge/gravity duality (or holography for short) may be better suited.
The main purpose of this letter is to investigate compatibility of state-of-the-art holographic models for cold and dense QCD matter with the hypothesis that the smaller companion in GW190814 was a rapidly rotating NS.
We also compare our results to the universal mass ratio $M_{\mathrm{max}}/M_{\mathrm{TOV}} = 1.203^{+0.022}_{-0.022}$ proposed in~\cite{Breu:2016ufb}.

An interesting prediction of the holographic model we are using (V-QCD model~\cite{Jarvinen:2011qe}) is the presence of a strong first order nuclear to quark matter phase transition.
Therefore in this letter we, in particular, analyze how the phase transition affects the results derived from the data for GW190814.
We remark that the V-QCD model is one of the very few models which is able to describe both the nuclear and quark matter phases and therefore the phase transition in a single framework.

The rest of the paper is structured as follows.
In Sec.~\ref{sec:2} we introduce the holographic EoSs we are studying.
In Sec.~\ref{sec:3} we discuss the model for rotating stars we are using and their stability.
In Sec.~\ref{sec:4} we present results for the NS properties and mass-radius curves.
Finally, in Sec.~\ref{sec:5} we summarize and conclude.
Unless stated otherwise we use units where $c=G=1$.

\section{Equation of State}\label{sec:2}

We follow an approach \cite{Ecker:2019xrw,Jokela:2020piw} where the strongly coupled  bulk of the NS is modeled by employing the gauge/gravity duality whereas a traditional field theory approach is used for the crust. 

The holographic V-QCD model is a fusion of improved holographic QCD~\cite{Gursoy:2007cb,Gursoy:2007er} for the glue dynamics and an approach to include quarks based on brane actions~\cite{Bigazzi:2005md,Casero:2007ae}.
This approach is effective: the model contains a large number of parameters that must be determined by comparing to known properties of QCD, e.g., lattice data for the thermodynamics of QCD at finite temperature.
We study three different versions of V-QCD, referred to as soft, intermediate, and stiff, obtained through different fits to lattice QCD data in~\cite{Jokela:2018ers}\footnote{The ``soft'' (``stiff'') variant is given by the fit 5b (8b) in~\cite{Jokela:2018ers}.
The intermediate EoS was obtained by interpolating between the soft and stiff EoSs, using the fit result 7a from~\cite{Jokela:2018ers} as a guiding point.}.
As an important constituent for the current study, nuclear matter was included in the model by using a simple approximation scheme in~\cite{Ishii:2019gta}.
This model has been employed previously in NS merger simulations \cite{Ecker:2019xrw}, and to study the properties of QCD and static NSs based solely on predictions for the quark matter phase~\cite{Jokela:2018ers,Chesler:2019osn,Hoyos:2020hmq} and including holographic nuclear matter~\cite{Jokela:2020piw}.

To model the crust we use the Akmal-Pandharipande-Ravenhall (APR) \cite{Akmal:1998cf} EoS up to number densities $n<1.6\,n_s$, where $n_s=0.16\,\mathrm{fm}^{-3}$ is the nuclear saturation density.
The point $n_\mathrm{tr} =1.6 n_s$, where we match the holographic model with the low density nuclear matter model, is our estimate for the density where the holographic approach becomes more reliable than the traditional (effective field theory) approach. 
NSs with high masses, which are the focus of this letter, are mostly sensitive to the high density regime described by holography in our approach.
This has been shown in~\cite{Jokela:2020piw} by analyzing a larger set of ``hybrid'' EoSs of this kind, with various other nuclear models for the crust (in addition to APR) and spanning a wide range of values of $n_\mathrm{tr}$. 
We will also verify this here by carrying out a scan of basic observables over such a set of EoSs. 
All EoSs in this work assume beta equilibrium and zero temperature. 

\begin{figure}[ht!]
 \includegraphics[width=0.98\linewidth]{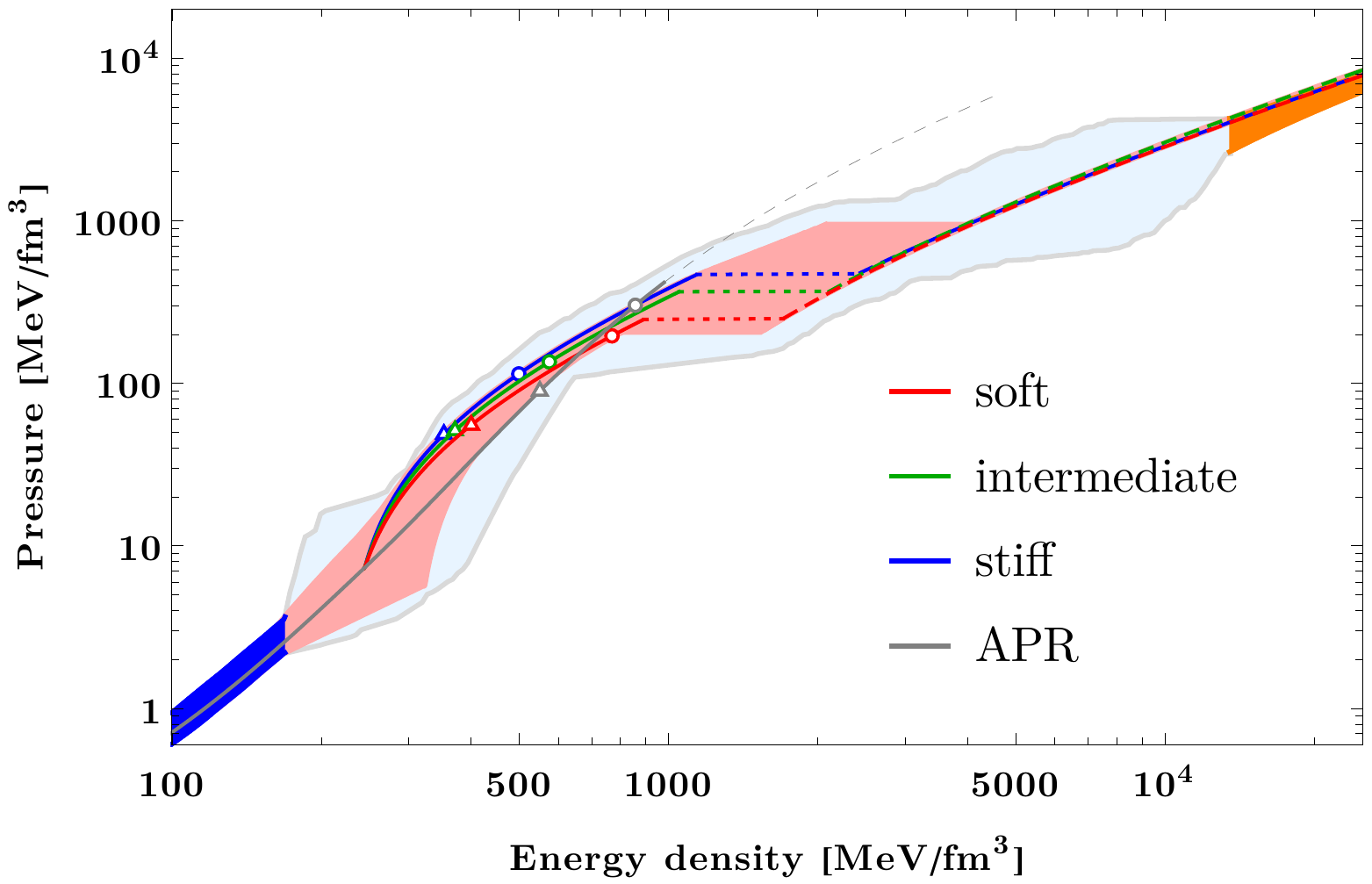}
 \caption{Equations of state.
 Red, green and blue curves are the V-QCD EoSs where first order phase transitions are indicated by dotted lines and the quark phases by dashed lines.
 Gray curve is the APR EoS whose superluminal part is the dashed part of the curve.
 Static $1.4\,M_\odot$ and $2\,M_\odot$ stars are marked by triangles and circles, respectively.
 Blue and orange bands indicate uncertainty in nuclear theory and perturbative QCD, respectively.
 The light red (light blue) band is spanned by holographic (general quadrutropic) interpolations between the low and high density limits.}
 \label{fig:eos}
\end{figure}

In Fig.~\ref{fig:eos} we show the three EoSs together with theoretical uncertainties in nuclear theory~\cite{Tews:2012fj} (blue band) and perturbative QCD~\cite{Kurkela:2009gj} (orange band) at low and high density, respectively, and bounds from causality and current astrophysical observations in between. 
We show two different bands at intermediate density: The light blue band is spanned by quadrutropic interpolations (following~\cite{Annala:2017llu}) between the low and high density results which also satisfy the astrophysical bounds: the maximal mass of static NSs is at least $2 M_\odot$ and the tidal deformability $\Lambda_{1.4}$ (at NS mass $M=1.4\,M_\odot$) is less than the  bound 580 obtained from the analysis of GW170817 by LIGO/Virgo (low-spin prior at 90\% confidence level)~\cite{TheLIGOScientific:2017qsa,Abbott:2018exr}.
The light red band is spanned by the aforementioned larger ensemble of V-QCD EoSs which satisfy the same bounds~\cite{Jokela:2020piw}.
Red, green and blue curves are the V-QCD EoSs which are matched at $\approx 244\,\mathrm{MeV}/\mathrm{fm}^3$ to the APR curve shown in gray.
Solid lines at low and intermediate densities correspond to the confined nuclear matter phase, dashed lines are in the deconfined quark matter phase and dotted horizontal lines are mixed phases between the two.
The thin dashed part of the gray curve is where the APR model has speed of sound larger than the speed of light and therefore clearly is outside its range of applicability. 

\section{Uniformly Rotating Stars}\label{sec:3}

We study properties of cold relativistic non-rotating and uniformly rotating stars modeled as stationary, ideal fluid distributions in general relativity using the publicly available RNS code~\cite{Cook:1993qr,Stergioulas:1994ea}.

To estimate $M_\mathrm{max}$ we construct sequences of uniformly rotating stars with different central number density $n_c$ and fixed angular momentum $J$.
These sequences are bounded from below by the so-called Keplerian or mass-shedding limit in which centrifugal and gravitational forces at the equator of the star exactly cancel.
The upper bound of these sequences is determined by black hole collapse which in our setup is either induced by a secular instability or the abrupt softening of the EoS at the phase transition.
The onset of the secular instability can be approximately determined with the turning-point criterion~\cite{1988ApJ...325..722F}
\begin{equation}\label{eq:rotcond}
\left.\frac{\partial M(n_c,J)}{\partial n_c}\right|_{J=\mathrm{const.}} =0\,.
\end{equation}
For non-rotating ($J=0$) stars \eqref{eq:rotcond} is necessary and sufficient to determine $M_{\mathrm{TOV}}$.
For rotating stars the turning-point criterion is only sufficient, but not necessary, i.e., rotating stars that are stable according to \eqref{eq:rotcond} can still be dynamically unstable and the onset of instability can appear at densities slightly lower than the turning-point density \cite{Takami:2011zc}.
In this work we neglect dynamical instabilities and approximate $M_\mathrm{max}$ by intersecting the Keplerian sequence either with lines constructed from \eqref{eq:rotcond} or with the number density at the phase transition $n_{\mathrm{QM}}$.
In the next section we show examples where both possibilities are realized, including a ``mixed'' case where $M_{\mathrm{TOV}}$ is determined by the phase transition and $M_{\mathrm{max}}$ by the secular instability.

\section{Results}\label{sec:4}

\begin{figure*}[ht!]
 \includegraphics[height=0.2\textheight]{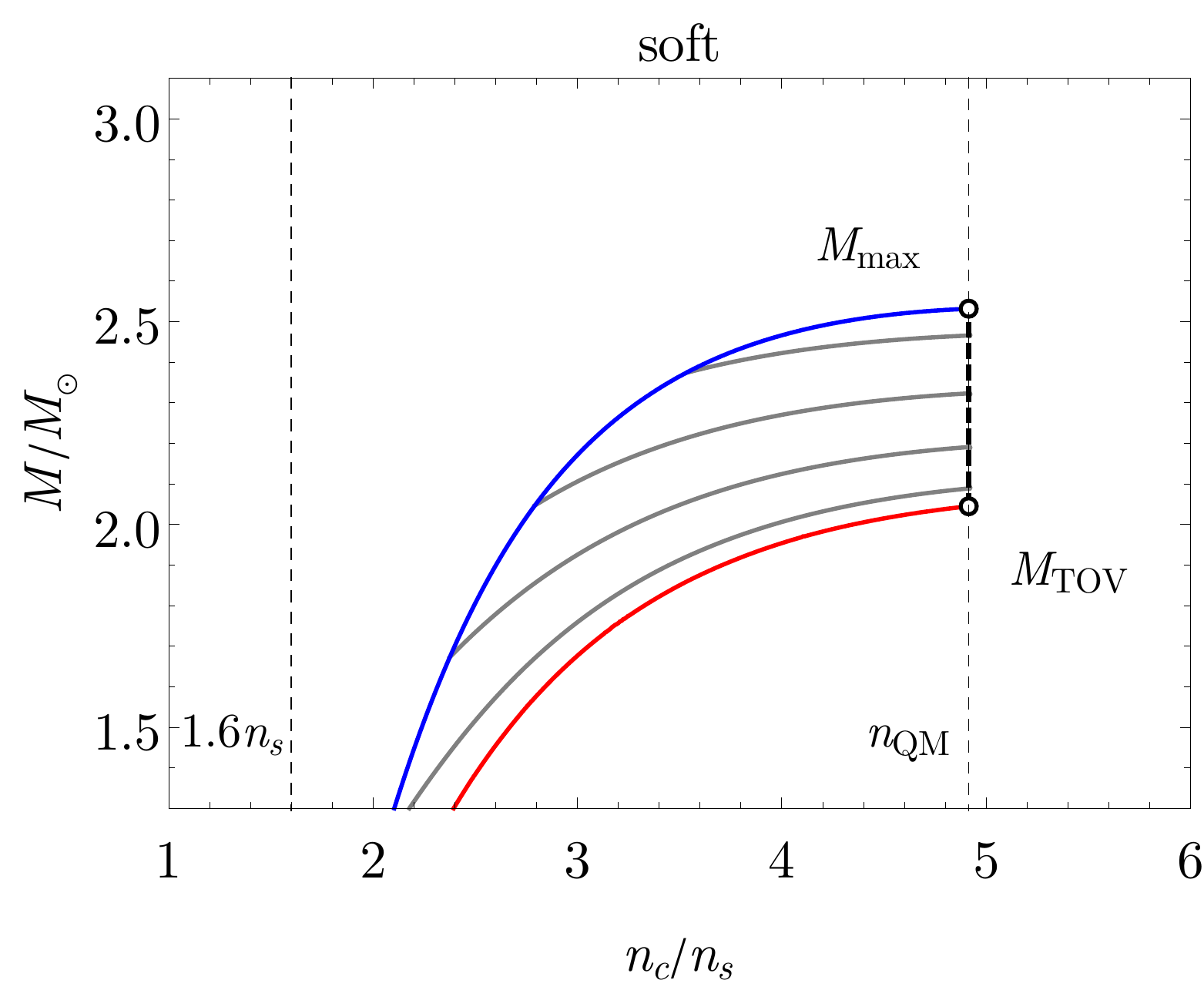}\quad\includegraphics[height=0.2\textheight]{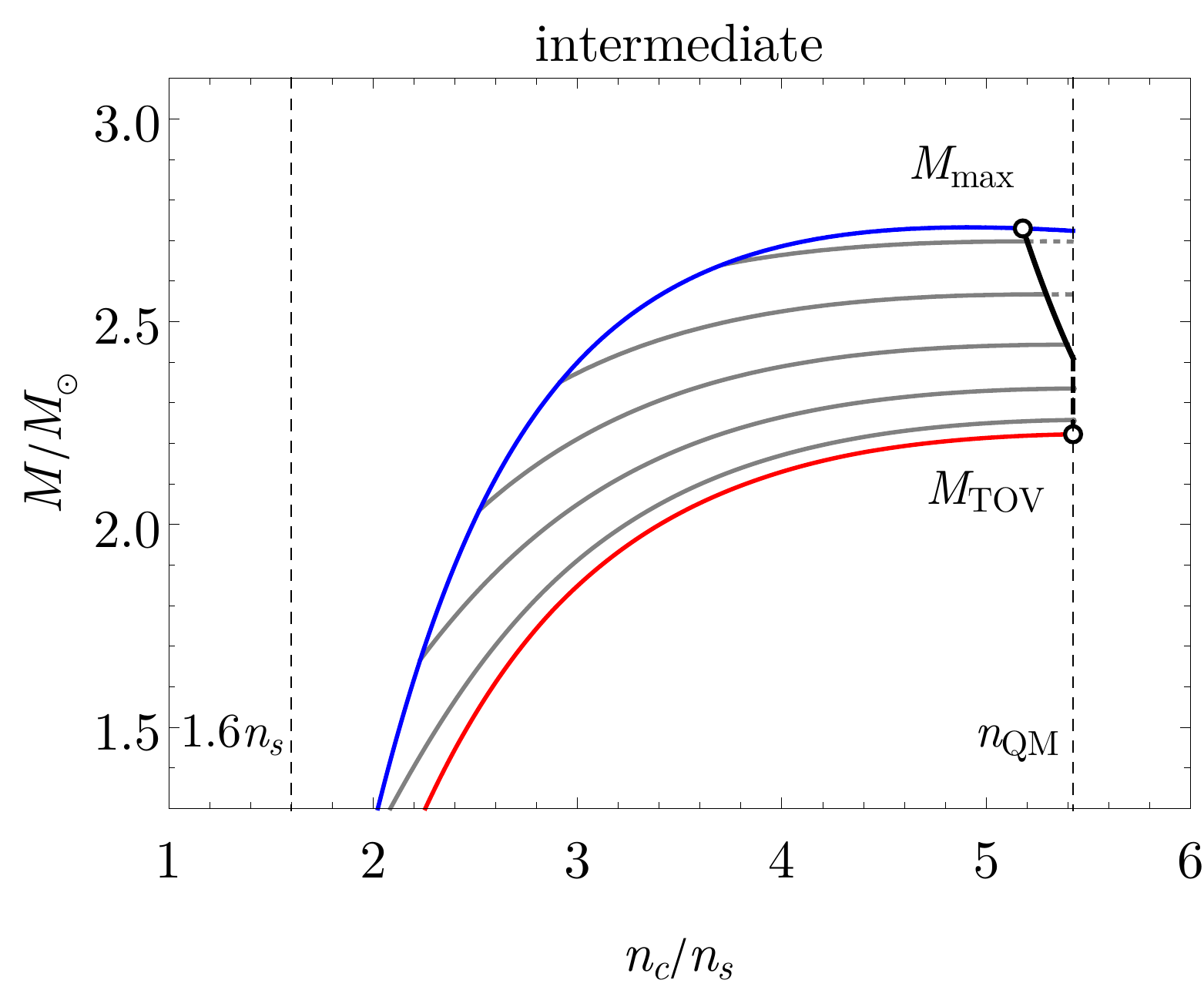}\quad\includegraphics[height=0.2\textheight]{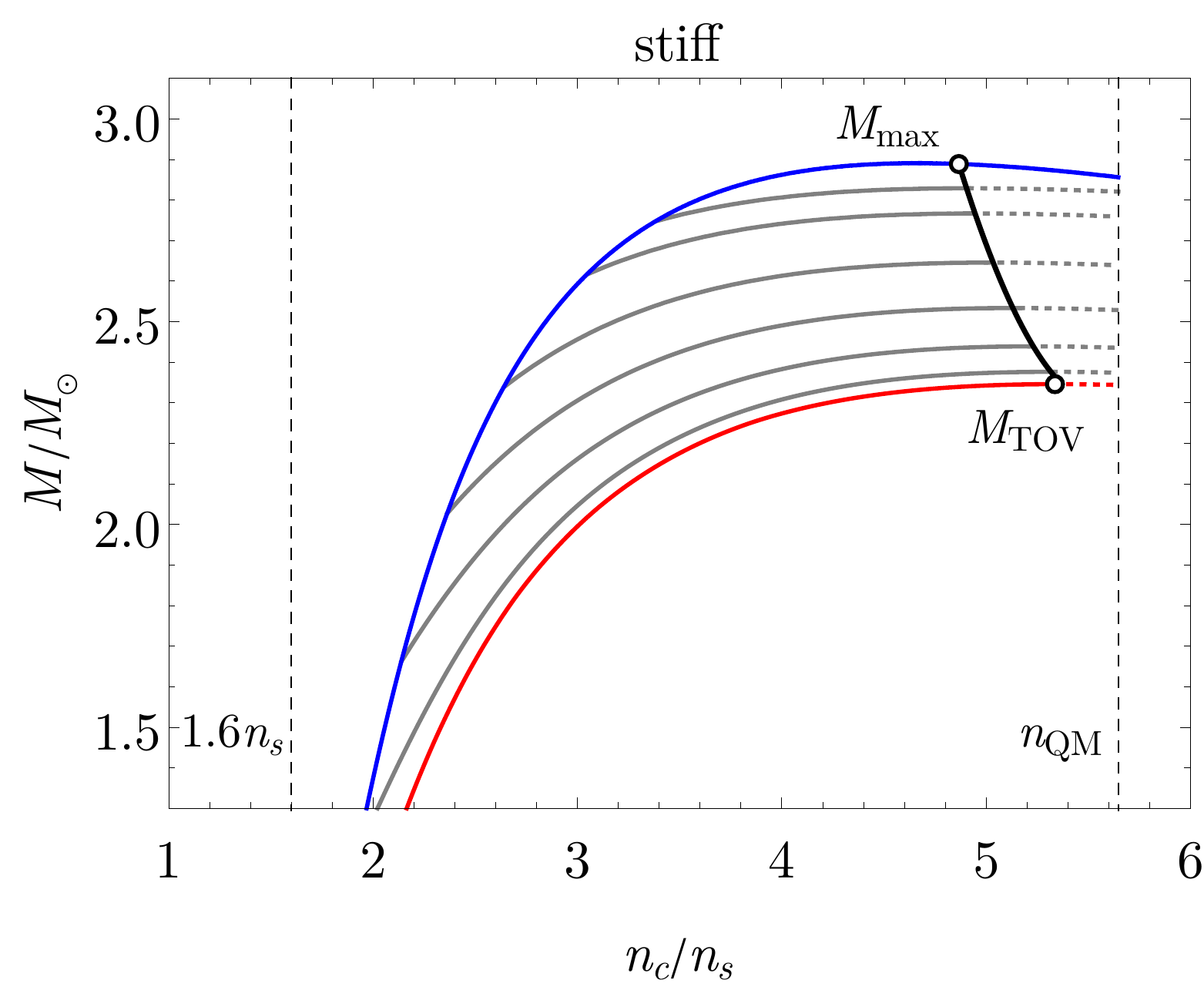}
 \caption{Mass as function of the central number density $n_c$ in units of the saturation density $n_s$ for soft (left), intermediate (center) and stiff (right) EoS. Red (blue) curves are for non-rotating (maximally rotating) configurations, vertical dashed lines indicate the matching density $1.6\,n_{s}$ and the density at the deconfinement phase transition $n_{QM}$.
 Gray lines are sequences with constant angular momentum.
 Black solid (dashed) curves mark the onset of the secular (phase transition induced) instability.}
 \label{fig:ME}
\end{figure*}

In Fig.~\ref{fig:ME} we show the masses of uniformly rotating NSs as function of the central baryon number density for the three variants of EoSs.
For the soft EoS (left plot) the onset of instability (black dashed line) is due to the abrupt softening of the EoS as the nuclear to quark matter transition is reached, independently of the angular momentum of the star.
In contrast to \cite{Paschalidis:2017qmb,Montana:2018bkb,Bozzola:2019tit} the holographic model predicts no stable twin or hybrid star solutions.

For the stiff EoS (right plot), however, the secular instability (black solid curve) is reached at densities below the transition density $n_{\mathrm{QM}}$.
The intermediate EoS shows the ``mixed'' behavior where the onset of instability is due to the phase transition (secular instability) for slowly (rapidly) rotating stars. 

Super-Massive Neutron Stars (SMNS), i.e. axisymmetrically rotating NSs with mass $M_{\mathrm{TOV}}\le M \le M_{\mathrm{max}}$, are expected to form in low and intermediate mass NS binary systems $M/M_{\mathrm{TOV}} \lesssim 1.5$ \cite{Baiotti:2016qnr}.
During its dynamic evolution a SMNS can lose angular momentum by a number of mechanisms
(electromagnetic emission, neutrino losses, etc.)
until it either ends up as stable, non-rotating NS or reaches a point of instability where it collapses into a black hole.
In the later case the lifetime of the star depends strongly on the high density part of the EoS where non-perturbative effects such as the deconfinement phase transition are crucial.
Our stars collapse because they reach the phase transition or the turning-point line.
By monotonicity of the turning-point line, it is sufficient to check if $M_\mathrm{TOV}$ is located at a turning-point to exclude the possibility of a phase transition induced collapse for all rotating NS, like it is for example the case in the stiff model shown in Fig.~\ref{fig:ME} (right).

\begin{figure}[ht!]
 \includegraphics[width=0.8\linewidth]{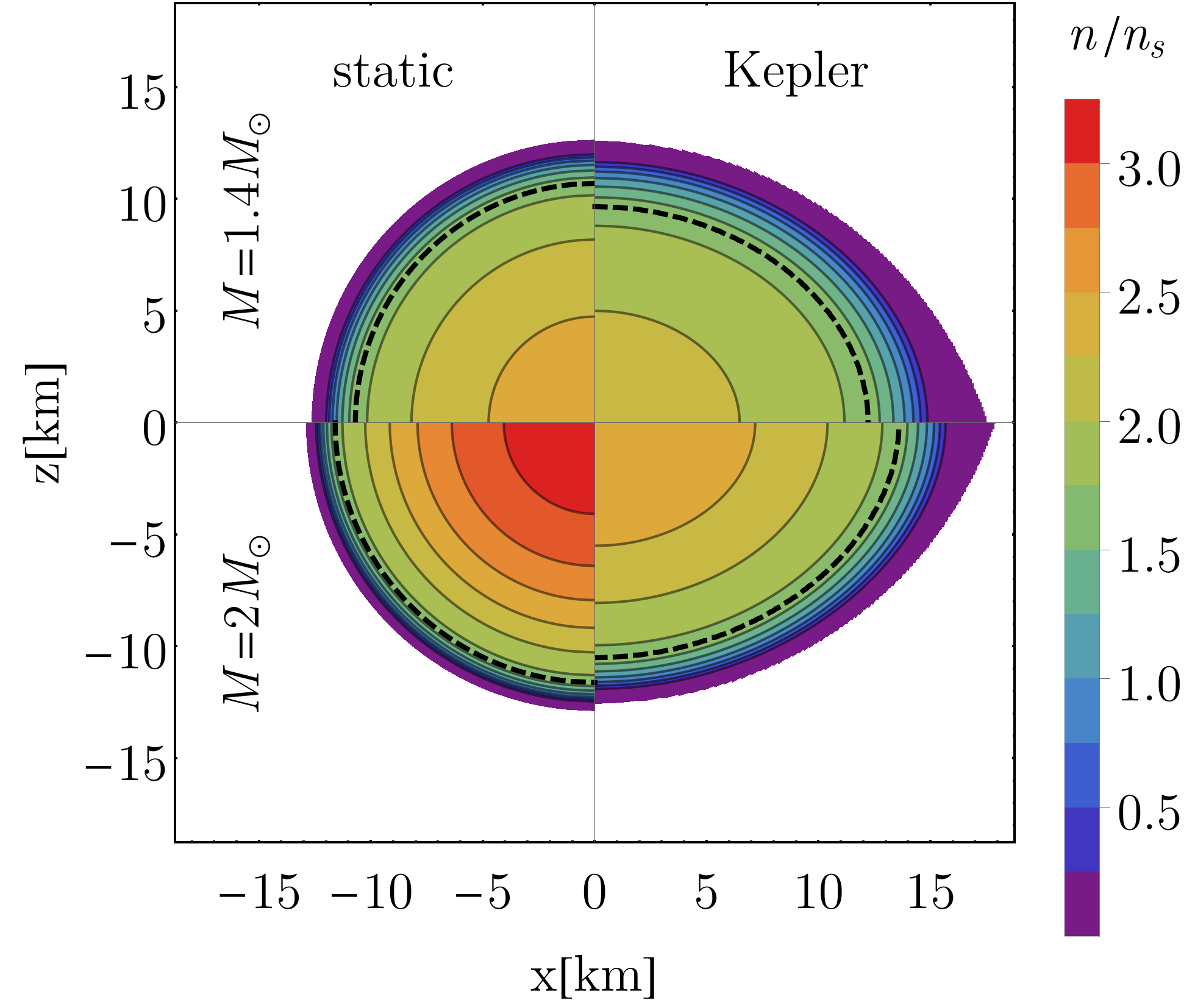}
 \caption{Density profile of static (left) and maximally rotating (right) stars with $1.4\,M_\odot$ (top) and $2\,M_\odot$ (bottom) for the stiff model. The black dashed line is the iso-contour at matching density $n=1.6\,n_s$ and in white regions $n/n_s<10^{-13}$.}
 \label{fig:star}
\end{figure}

The quadrants in Fig.~\ref{fig:star} show the number density profile of a static star (left) and a star, spinning around the $z$-axis at Keplerian, mass-shedding frequency $f_{\mathrm{max}}$ (right) for two choices for the mass.
Notice that almost all matter in the cores of the stars is described by the holographic model; the dashed black line indicates the $n=1.6\,n_s$ iso-number density surface at which APR transitions to the V-QCD EoS. 
In the non-rotating configuration with $M =1.4\,M_\odot$ ($M =2\,M_\odot$) $R_{\mathrm{match}}/R_e=0.85$ ($R_{\mathrm{match}}/R_e=0.90$), where $R_e$ and $R_{\mathrm{match}}$ are the maximal radius and the radius at matching density in the equatorial plane, respectively.

\begin{figure*}[ht!]
 \includegraphics[height=0.20\textheight]{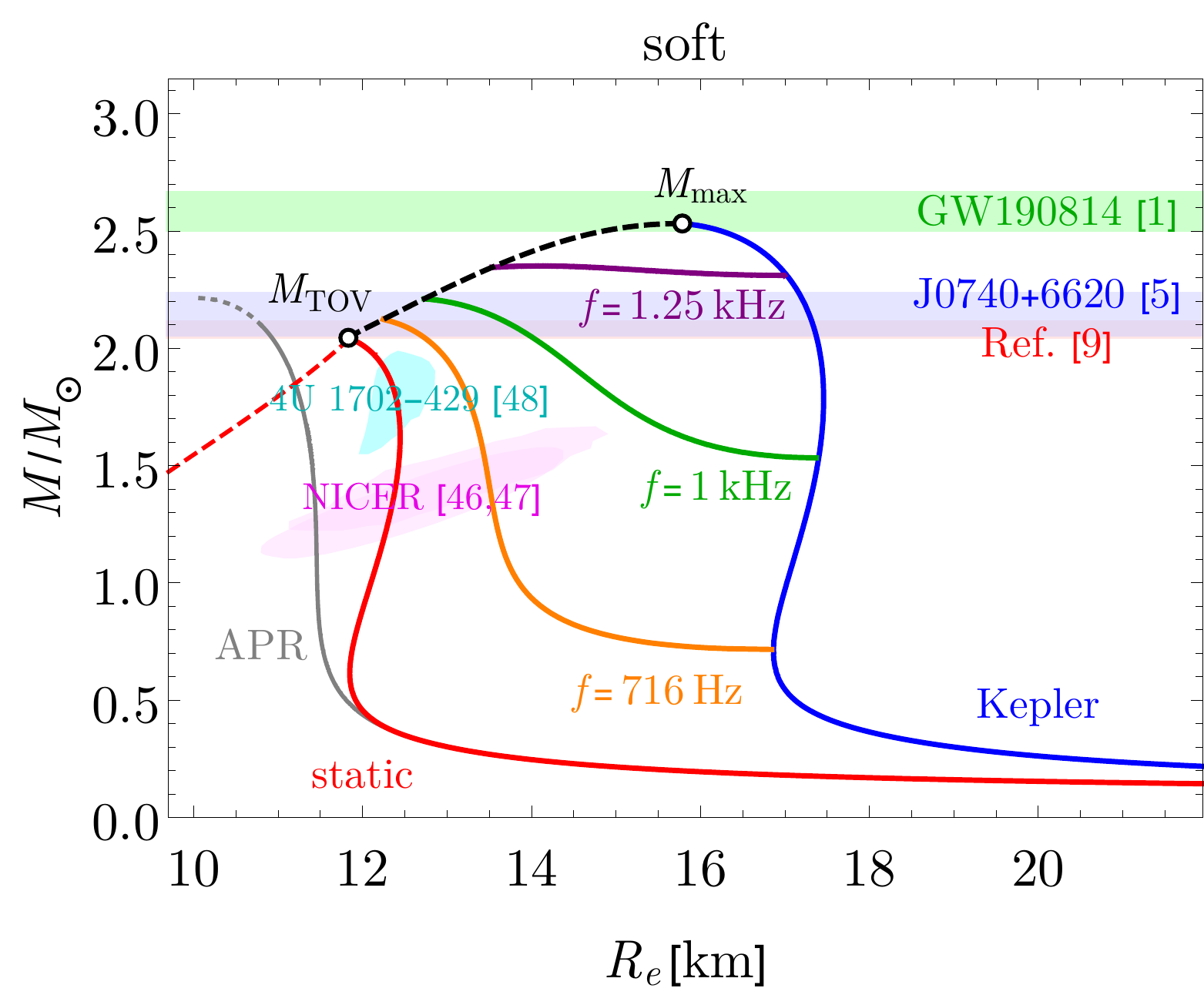}\quad\includegraphics[height=0.20\textheight]{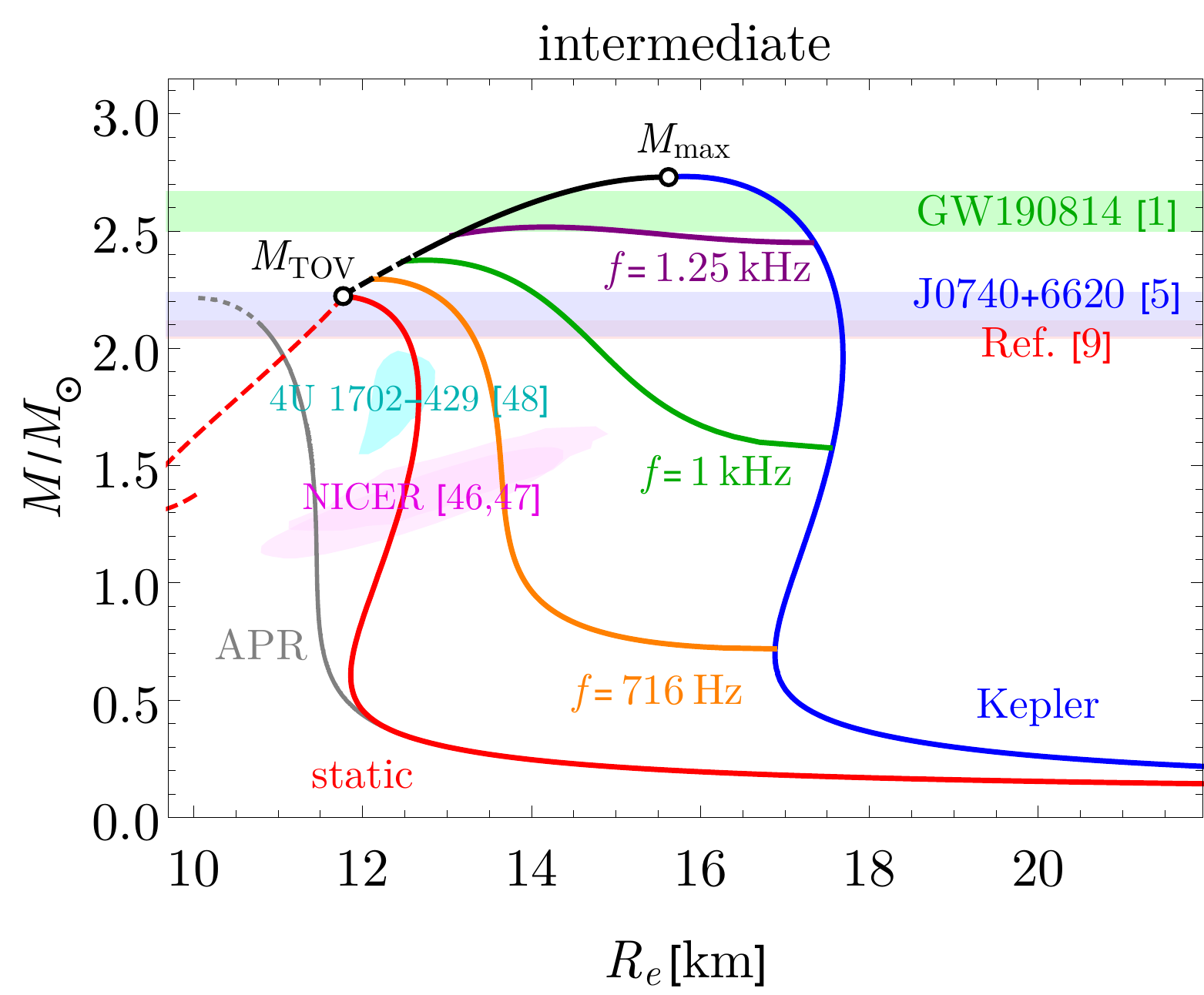}\quad\includegraphics[height=0.20\textheight]{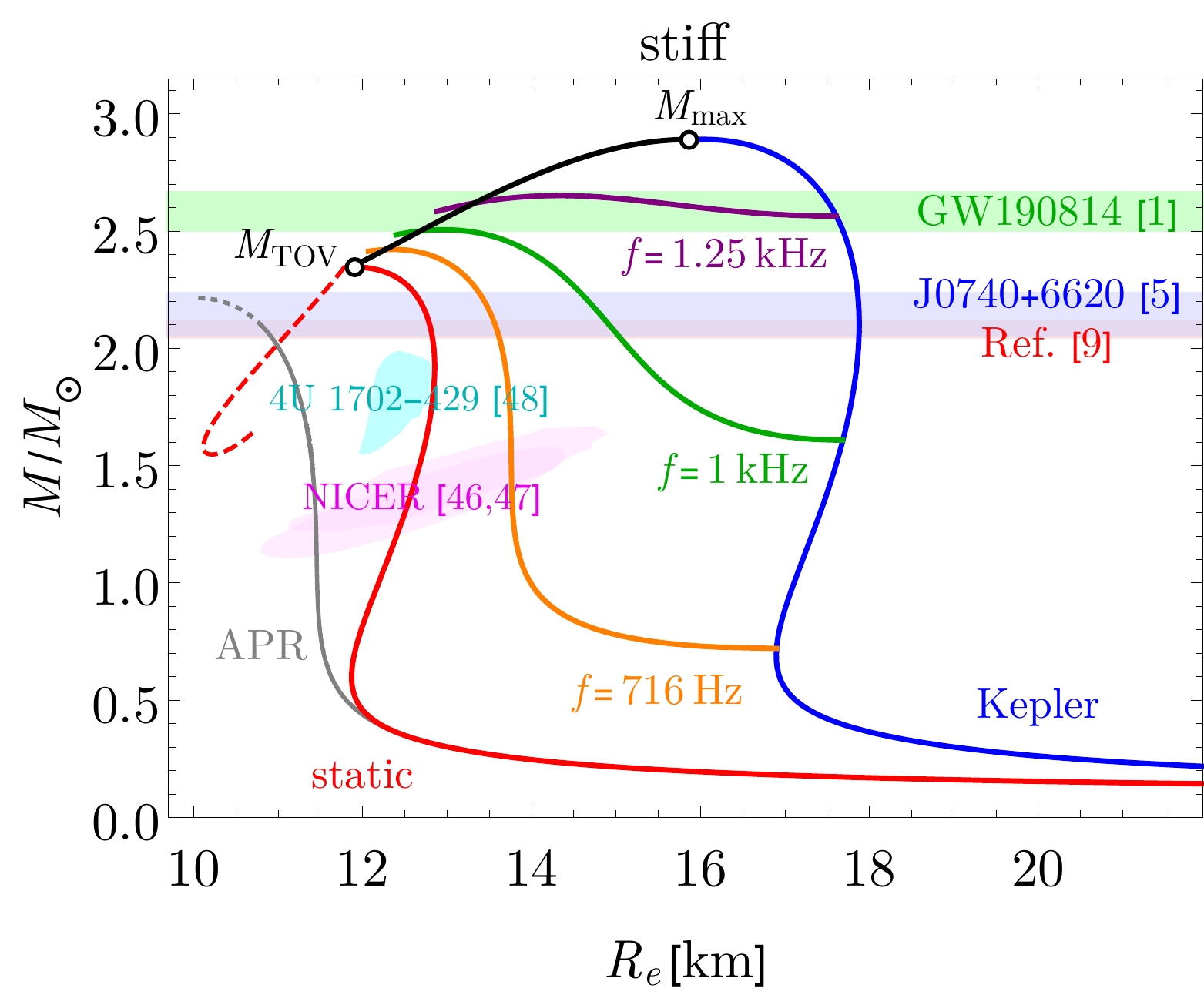}
 \caption{Mass-Radius relations for soft (left), intermediate (center) and stiff (right) EoS.
 Red (blue) curves are for non-rotating (maximally rotating) configurations, the red dashed line is the unstable branch in the quark matter phase, green, purple and orange lines are sequences of fixed rotational frequency.
 The horizontal bands show the result for the mass of the secondary component of GW190814 (light green) and the most stringent bounds on the maximum mass of static NSs (light blue and light red)~\cite{Cromartie:2019kug,Most:2020bba}.
 In addition we show observational bounds deduced from PSR J0030+0451 by NICER~\cite{Riley:2019yda,Miller:2019cac} (pink ellipses) and from the measurement of the X-ray binary 4U 1702-429~\cite{Nattila:2017wtj} (cyan area).}
 \label{fig:MR}
\end{figure*}

In Fig.~\ref{fig:MR} we show mass-radius relations for the three holographic EoSs and compare them to the results from GW190814 and a selection of other measurements of NS masses and radii.  
The static stars in the soft model (solid red curve in the right plot) reach the maximum mass $M_\mathrm{TOV}=2.04 M_\odot$, a value only barely consistent with direct mass measurements~\footnote{We show only the result from~\cite{Cromartie:2019kug}.
At 1-$\sigma$ level the other measurements mentioned in the introduction do not affect the constraint on the maximum mass significantly.} and the bound derived in~\cite{Most:2020bba} based on the GW190814 event, shown as light blue and light red bands, respectively. 
We also notice that all stable stars consistent with the GW190814 measurement (i.e., within the light green band) are almost maximally rotating.
The other models (intermediate and stiff) satisfy these bounds easily, but even for the stiff model high rotation frequencies $\gtrsim$~1~kHz are required to reach the green band.
This is well above the frequency $f=716\,\mathrm{Hz}$ of the fastest spinning pulsar observed so far, PSR J1748-2446ad~\cite{Hessels:2006ze}. 

There are also estimates for the radii of NSs using the X-ray channel.
We show the results from the measurement of PSR~J0030+0451 by the NICER collaboration~\cite{Riley:2019yda,Miller:2019cac} as well as the measurements of the low-mass X-ray binary 4U 1702-429 obtained by the Rossi X-Ray Timing Explorer~\cite{Nattila:2017wtj}.
Our results for slowly rotating NSs agree well with these results for all three models.

\begin{figure}[ht!]
 \includegraphics[width=\linewidth]{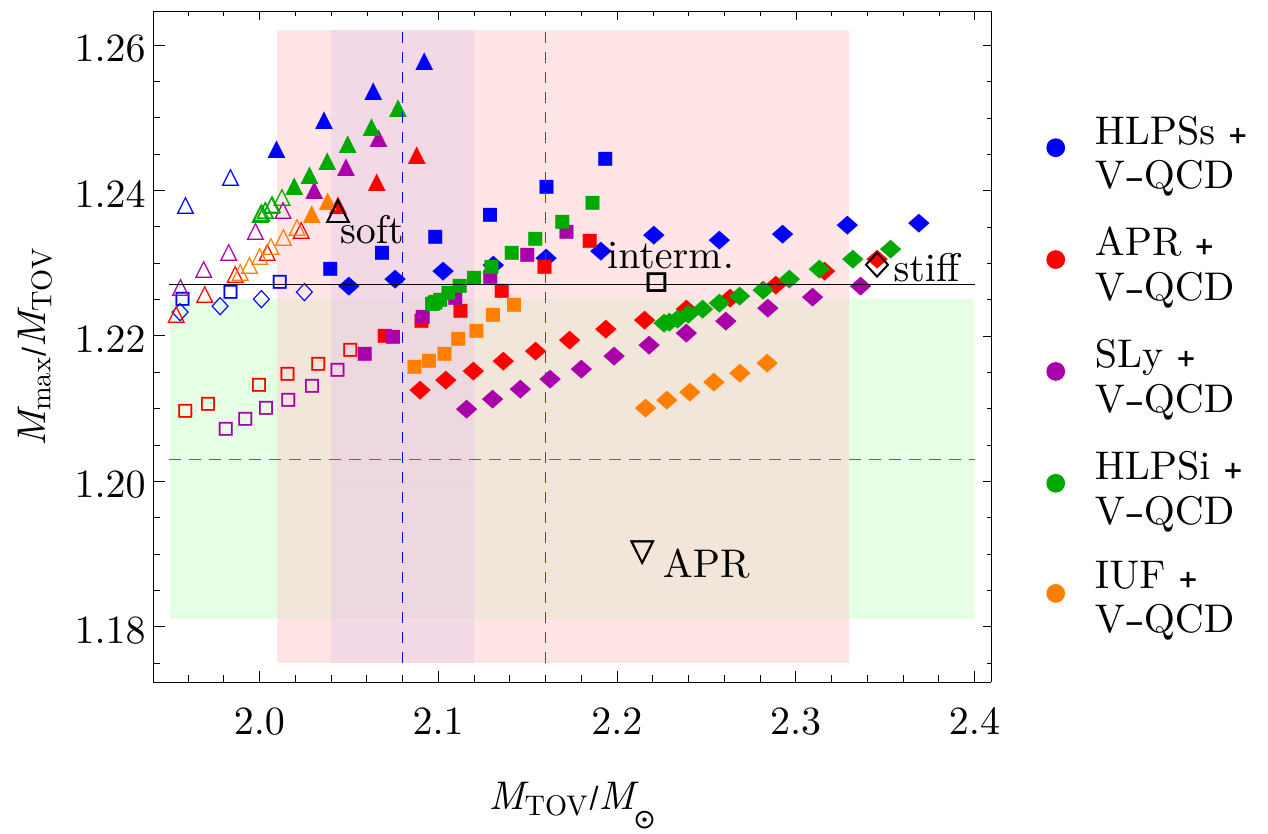}
 \caption{The mass ratio $M_\mathrm{max}/M_\mathrm{TOV}$ vs. $M_\mathrm{TOV}$ for an ensemble of hybrid EoSs from~\cite{Jokela:2020piw}. The colored markers show the results for different low density nuclear matter models, indicated in the legend, as the matching density $n_\mathrm{tr}$ varies. Triangles, squares, and diamonds use soft, intermediate, and stiff versions of the holographic model, respectively. The filled (open) markers are EoSs for which the Keplerian curve reaches (fails to reach) the GW190814 band in Fig.~\protect\ref{fig:MR}. The EoSs used in the other figures are shown with large black open markers. The bands green, blue, and red bands show the estimates for the mass ratio~\cite{Breu:2016ufb}, minimum of $M_\mathrm{TOV}$ based on GW190814~\cite{Most:2020bba}, and maximum of $M_\mathrm{TOV}$~\cite{Rezzolla:2017aly}, respectively. The horizontal black line shows our fit result from~\protect\eqref{eq:massratiofit}.}
 \label{fig:massratio}
\end{figure}

In addition to the three holographic EoSs of Fig.~\ref{fig:eos}, we have carried out a scan over all EoSs constructed in~\cite{Jokela:2020piw} which span the light red band in Fig.~\ref{fig:eos}.
Apart from APR, these additional EoSs use the following nuclear matter models at low density: the soft and intermediate variants of the Hebeler-Lattimer-Pethick-Schwenk (HLPS)~\cite{Hebeler:2013nza}, Skyrme Lyon (SLy)~\cite{Haensel:1993zw,Douchin:2001sv}, and IUF~\cite{Hempel:2009mc,Fattoyev:2010mx}.
We also allowed the matching density $n_\mathrm{tr}$ to vary within the range from $1.3 n_s$ to $2.2n_s$ and required that the EoSs comply with the LIGO/Virgo bound $\Lambda_{1.4}<580$~\cite{Abbott:2018exr}.
Our results for the distribution of the mass ratio $M_\mathrm{max}/M_\mathrm{TOV}$ and the maximum non-rotating mass $M_\mathrm{TOV}$ are shown in Fig.~\ref{fig:massratio}.
The mass ratios are mildly shifted upwards with respect to the fit $M_\mathrm{max}/M_\mathrm{TOV} =1.203 \pm 0.022$ for EoSs without a deconfinement transition (green dashed line and band)~\cite{Breu:2016ufb}.
We stress that this happens even though our NSs are fully hadronic; for models admitting hybrid stars with quark matter cores, larger deviations are possible~\cite{Bozzola:2019tit}.
Consequently, some of the EoSs with the soft variant of V-QCD lie slightly below the estimate of the lowest possible $M_\mathrm{TOV}$ (blue dashed line and band) from~\cite{Most:2020bba}.
Moreover we note that the stiffest holographic EoSs are able to produce stable NSs a bit above the bound  $M_{\mathrm{TOV}}<2.16^{+0.17}_{-0.15}M_\odot$ of~\cite{Rezzolla:2017aly} (red dashed line and band) and the estimate $M_{\mathrm{TOV}} \lesssim 2.3 M_\odot$ of~\cite{Shibata:2019ctb}. 

In order to compare directly to the result of~\cite{Breu:2016ufb}, we fitted the dependence of $M_\mathrm{crit}$ on the scaled angular momentum $j=J/M_\mathrm{crit}^2$, where $M_\mathrm{crit}$ is the value of the mass at the onset of instability, given as the black curves in Fig.~\ref{fig:ME}.
We used the data for all EoSs shown with filled colored markers in Fig.~\ref{fig:massratio}.
The fit to the formula~\cite{Breu:2016ufb}
\begin{equation}
\frac{M_\mathrm{crit}}{M_\mathrm{TOV}} = 1 +a_2 \left(\frac{j}{j_\mathrm{Kep}}\right)^2+a_4 \left(\frac{j}{j_\mathrm{Kep}}\right)^4 \,, 
\end{equation}
where $j_\mathrm{Kep}$ is the value at the intersection with the Keplerian curve, gives $a_2 = 0.1603$ and $a_4 = 0.0667$. 
Evaluating this fit at $j=j_\mathrm{Kep}$ we obtain the estimate
\begin{equation} \label{eq:massratiofit}
    \frac{M_\mathrm{max}}{M_\mathrm{TOV}} = 1.227^{+0.031}_{-0.016} \,,
\end{equation}
where the error bars indicate the largest deviation from the fit.

In addition we estimate the average ratio of the maximum rest mass $M_{\mathrm{b,max}}$ and $M_{\mathrm{max}}$ for all our EoSs
\begin{equation} \label{eq:massBratio}
    \frac{M_\mathrm{b,max}}{M_\mathrm{max}} = 1.177^{+0.018}_{-0.020} \,, 
\end{equation}
where the error bars again indicate the largest deviation from the central value.
Also here our value is slightly above the value $1.171^{+0.014}_{-0.014}$ (two sigma level) obtained in \cite{Rezzolla:2017aly} for EoSs without phase transition.

Table \ref{tab:NSprop} summarizes salient features of rotating and non-rotating stars, such as number density at the deconfinement phase transition ($n_\mathrm{QM}$), maximum mass of non-spinning ($M_\mathrm{TOV}$) and spinning ($M_\mathrm{max}$) stars, the maximum rest mass ($M_{\mathrm{b,max}}$) of spinning stars, range of equatorial radii ($R_{\mathrm{e}}$) and tidal deformability ($\Lambda_{1.4}$) of $1.4\,M_\odot$ stars and maximum rotation frequency ($f_{\mathrm{max}}$).
\begin{table}[h!tb]
 \caption{Properties of static and rotating stars.}
 \begin{tabular}{ccccccccc}\hline\hline
  Model & $\frac{n_\mathrm{QM}}{n_s}$ & $\frac{M_\mathrm{TOV}}{M_\odot}$  & $\frac{M_\mathrm{max}}{M_\mathrm{TOV}}$& 
          $\frac{M_\mathrm{b,max}}{M_\mathrm{max}}$ &
          $\frac{R_{\mathrm{e},1.4}}{\mathrm{km}}$ & $\Lambda_{1.4}$ & $\frac{f_\mathrm{max}}{\mathrm{kHz}}$  
  \\
  \hline\hline
  soft & $4.89$ & $2.04$ & $1.238$ & $1.172$ & $[12.38,17.33]$  & $493$ & $1.45$  
  \\
  interm. & $5.43$ & $2.22$ & $1.228$ & $1.186$ & $[12.51,17.44]$ & $536$ & $1.54$ 
  \\
  stiff & $5.61$ & $2.35$ & $1.231$ & $1.194$ & $[12.60,17.52]$ & $567$ & $1.60$  
  \\
  \hline
  APR & --    & $2.21$ & $1.192$ & $1.202$ & $[11.40,16.14]$ & $260$ & $2.01$ 
  \\
  \hline
\end{tabular}
\label{tab:NSprop} 
\end{table}

\section{Conclusion}\label{sec:5}

In this letter, we analyzed spinning hadronic NSs with EoSs having a deconfinement transition from nuclear to quark matter.
The analysis was made possible by using state-of-the-art holographic models for dense QCD which include both the nuclear and quark matter phases (V-QCD), and therefore give controlled predictions for the properties of the phase transition.
The phase transition is strongly first order as the EoS in the NM (QM) phase is relatively stiff (soft).
Interestingly, this picture is similar to what arises in another non-perturbative approach, i.e., the functional renormalization group method~\cite{Drews:2016wpi,Otto:2019zjy}.

Apart from the phase transition, the holographic models predict that the nuclear matter EoSs is relatively stiff at high densities, making it easy to reach high masses for both non-rotating and rotating NSs.
For stiff variants of the model, we find that the maximal masses of non-rotating (rotating) stars are around 2.35 (2.9) solar masses.
Therefore these models are easily consistent with the interpretation~\cite{Most:2020bba,Dexheimer:2020rlp} that the secondary component of the binary merger event GW190814 is a rapidly spinning NS.
However, even for the stiffest EoSs the frequencies required for this interpretation are high: we find that $f \gtrsim$~1~kHz, which is close to the mass shedding limit ($\sim$ 1.5~kHz) and  clearly above the fastest pulsar rotation frequency observed so far, 716~Hz~\cite{Hessels:2006ze}.

Curiously, in the three models considered in this work the maximal masses of rotating and non-rotating stars are determined in different ways.
In the soft model, both $M_\mathrm{TOV}$ and $M_\mathrm{max}$, are determined by the phase transition.
In the intermediate model, $M_\mathrm{TOV}$ is determined by the phase transition and $M_\mathrm{max}$ of rapidly rotating stars by the secular instability.
In the stiff model, both $M_\mathrm{TOV}$ and $M_\mathrm{max}$ are determined by the secular instability.
We find it is sufficient to check if $M_\mathrm{TOV}$ is located at a turning-point to exclude the possibility of a phase transition induced collapse for all rotating NS for a given EoS.
Consequently, SMNSs formed in binary NS mergers with EoS satisfying this simple criterion will not produce any signatures of the phase transition in their gravitational wave signal.

Interestingly, we find the ratio of maximal masses of rotating vs. non-rotating models to be relatively high compared to models without deconfinement transition~\cite{Breu:2016ufb}; our fit result $M_{\mathrm{max}}/M_{\mathrm{TOV}}=1.227^{+0.031}_{-0.016}$ is shifted upwards with respect to their result, $1.203^{+0.022}_{-0.022}$.
This feature is pronounced for EoSs with the soft variant, which can reach $M_{\mathrm{max}}/M_{\mathrm{TOV}}\approx 1.26$.
Notice that it is indeed the soft variants where both $M_\mathrm{max}$ and $M_\mathrm{TOV}$ are determined by the instability induced by the phase transition, which supports the interpretation that the large values for the ratio arise due to the transition.

In this letter, we neglected some well known effects due to computational simplicity.
First, we used the turning-point criterion~\eqref{eq:rotcond} to estimate the onset of the secular instability.
To improve on this, we would need to study the impact of dynamical instabilities on our results by carrying out 3+1 dimensional simulations.
We expect, however, that carrying out the full analysis would change our results very little, because the mass of the NS is insensitive to the exact value of the critical central density near the onset of the instability (see, e.g.,~\cite{Takami:2011zc}).
Second, we studied rigidly rotating stars only.
It would be interesting to generalize our work to differentially rotating stars for which a similar universal ratio $M_{\mathrm{max,dr}}/M_{\mathrm{TOV}}=1.54^{+0.05}_{-0.05}$ has been proposed \cite{Weih:2017mcw}. 
We hope to return to this topic in future work.

\begin{acknowledgments}

We thank Niko Jokela, Antonios Nathanail, Luciano Rezzolla and Lukas Weih for useful discussions and comments on the manuscript.
TD was partially supported by the Israel  Science Foundation (ISF) grant \#1635/16 and the BSF grants  \#2015626 and \#2018722.
The work of MJ was supported in part by a center of excellence sup-ported by the ISF grant \#2289/18.
This research was also supported by an appointment to the JRG Program at the APCTP through the Science and Technology Promotion Fund and Lottery Fund of the Korean Government.
In addition, the research was supported by the Korean Local Governments -- Gyeongsangbuk-do Province and Pohang City.
\end{acknowledgments}

\bibliographystyle{apsrev4-1}
\bibliography{references}

\end{document}